\documentclass[sigconf]{acmart}

\AtBeginDocument{%
  }

\setcopyright{acmlicensed}
\copyrightyear{2018}
\acmYear{2018}
\acmDOI{XXXXXXX.XXXXXXX}
\acmConference[Conference acronym 'XX]{Make sure to enter the correct
  conference title from your rights confirmation email}{June 03--05,
  2018}{Woodstock, NY}
\acmISBN{978-1-4503-XXXX-X/2018/06}

\newcommand\yawei[1]{\textbf{\textcolor{blue}{YW: #1}}	}

\newcommand{\one}{\textbf{({\em i}\/)}\xspace}
\newcommand{\two}{\textbf{({\em ii}\/)}\xspace}
\newcommand{\three}{\textbf{({\em iii}\/)}\xspace}

\newcommand{\pb}[1]{\vspace{0.75ex}\noindent{\bf \em #1}}
\newcommand{\spb}[1]{\vspace{0.75ex}\noindent{\em \uline{#1}}}
\usepackage[normalem]{ulem}
\usepackage{tablefootnote}
\usepackage{xspace}
\usepackage{csquotes}
\usepackage{tcolorbox}
\usepackage{subcaption}
\usepackage{soul}
\usepackage{makecell}
\usepackage{mwe}
\usepackage{multirow}
\usepackage{graphicx}
\usepackage{nicematrix}
\usepackage{fontawesome}
\usepackage{tabularx}
\usepackage{booktabs}
\usepackage{pifont}
\usepackage{multirow}
\usepackage{amsmath}
\usepackage{svg}
\usepackage{verbatim}
\usepackage{xcolor}
\usepackage{xstring}
\usepackage{enumitem}
\usepackage{array}
\usepackage{diagbox}
\usepackage{textcomp}
\usepackage{bm}
\usepackage{caption}

\newcolumntype{L}[1]{>{\raggedright\let\newline\\\arraybackslash\hspace{0pt}}m{#1}}
\newcolumntype{C}[1]{>{\centering\let\newline\\\arraybackslash\hspace{0pt}}m{#1}}
\newcolumntype{R}[1]{>{\raggedleft\let\newline\\\arraybackslash\hspace{0pt}}m{#1}}

\newtcolorbox{takehome}{
    colback=gray!5,      
    colframe=black!75,   
}

\acmSubmissionID{310}

\begin{document}

\title[Toward Site-Aware MR Art Exhibitions]{Toward Site-Aware MR Art Exhibitions: A SLAM-Based Deployment Pipeline for Spatial Coherence and Exhibition Experience}


\author{Yawei Zhao}
\email{yzhao099@connect.hkust-gz.edu.cn}
\orcid{0000-0001-9298-0955}
\affiliation{%
  \institution{The Hong Kong University of Science and Technology (Guangzhou)}
  \city{Guangzhou}
  \country{China}
}

\author{Yiming Zhu}
\email{yzhucd@connect.ust.hk}
\orcid{0000-0002-7537-5582}
\affiliation{%
  \institution{Hong Kong University of Science and Technology}
  \department{IIP (AI)}
  \city{Hong Kong}
  \country{China}
}

\author{Hao Li}
  \email{hli307@connect.hkust-gz.edu.cn}
  \orcid{0009-0007-2434-4325}
\affiliation{%
  \institution{The Hong Kong University of Science and Technology (Guangzhou)}
  \city{Guangzhou}
  \country{China}
}

\author{Yuqi Liang}
    \email{yliang379@connect.hkust-gz.edu.cn}
    \orcid{0009-0003-2640-8311}
\affiliation{%
  \institution{The Hong Kong University of Science and Technology (Guangzhou)}
  \city{Guangzhou}
  \country{China}}

\author{Ao Yu}
    \email{aoyu@hkust-gz.edu.cn}
    \orcid{0000-0002-1520-1097}
\affiliation{%
  \institution{The Hong Kong University of Science and Technology (Guangzhou)}
  \city{Guangzhou}
  \country{China}}

\author{Anca-Simona Horvath}
 \email{ancahorvath@hkust-gz.edu.cn}
 \orcid{0000-0001-5371-5657}
\affiliation{%
 \institution{The Hong Kong University of Science and Technology (Guangzhou)}
 \city{Guangzhou}
 \country{China}
 }
 \affiliation{%
 \institution{Aalborg University}
 \city{Aalborg}
 \country{Denmark}
}

\author{Pan Hui}
\email{panhui@ust.hk}
\orcid{0000-0001-6026-1083}
\authornotemark[1]
\affiliation{%
  \institution{The Hong Kong University of Science and Technology (Guangzhou); The Hong Kong University of Science and Technology}
  \city{Guangzhou}
  \country{China}
}

\renewcommand{\shortauthors}{Yawei Zhao et al.}

\begin{abstract}

Mixed Reality (MR) is increasingly being used in exhibition settings to bring digital artworks into relation with the physical environment. However, existing MR exhibition systems are often confined to prototypes or case-specific deployments, offering limited guidance for large-scale practical implementation. To address this gap, this paper presents a practical pipeline for designing and deploying large-scale MR art exhibitions, treating spatial alignment not only as a technical mechanism but also as an experiential design decision. We first conducted a pilot study comparing marker-based and Simultaneous Localization and Mapping (SLAM)-based alignment methods in an MR exhibition setting. Based on the results, we developed a SLAM-based pipeline for MR exhibitions that integrates technical deployment with exhibition curation. We then evaluated the pipeline through both system overhead measures and users’ experiential feedback. The results show that spatial alignment influences not only technical stability, but also overall exhibition coherence, visitors’ sense of continuity and immersion, and artwork interpretation. These findings provide an empirically grounded reference for future large-scale MR art exhibition deployment.
\end{abstract}





\maketitle



\section{Introduction}


Mixed Reality (MR) technologies are increasingly used in cultural and exhibition settings such as museums, galleries, and public installations to support immersive and spatially situated experiences \cite{sanfilippo2025GamifyingCultural,trunfio2022MixedReality,tran2024SurveyMeasuring}. By overlaying digital elements onto the physical environment, MR allows artistic and cultural content to be experienced directly within real-world spaces, so that virtual and physical elements can be encountered together in context \cite{speicher2019WhatMixed,bekele2018SurveyAugmented,damala2008BridgingGap,bekele2021InfluenceCollaborative}.

Prior work has shown that MR systems can enrich exhibitions by facilitating visitor engagement, narrative interpretation, and spatial interaction with digital content \cite{hammady2020AmbientInformation,hammady2021InteractiveMixed}. A key technical requirement in these systems is stable spatial alignment, which refers to keeping virtual objects correctly matched to the physical environment in both position and orientation \cite{reviewmixed2020, mcgill2020quest, radanovic2023aligning}. This is essential for maintaining coherent overlays and makes tracking and registration critical technical concerns \cite{bekele2018SurveyAugmented,sanfilippo2025GamifyingCultural}. Spatial alignment is typically achieved through either marker-based or markerless methods \cite{reviewmixed2020}. In practice, marker-based approaches are often adopted because of their simplicity and ease of deployment, whereas markerless approaches using Simultaneous Localization and Mapping (SLAM) can support more flexible spatial alignment in larger-scale environments \cite{bekele2018SurveyAugmented}. Nonetheless, there was a limited understanding of how different spatial alignment approaches shape exhibition coherence and visitor experience in practice. Meanwhile, some existing studies have presented deployment demo systems or prototypes\cite{augmentedreality2025,hyperrealitycase2023,interactionblueprint,reconstructionvisual2025}. However, these efforts are often limited in specific installations, prototype demonstrations, or case studies. There remains a lack of generalizable and empirically evaluated pipelines for large-scale MR art exhibition deployment.

To bridge this gap, this paper explores three related questions concerning the design and deployment of MR art exhibitions. First, we examine 
\textit{what role do commonly used spatial alignment approaches play in MR exhibition settings, particularly in shaping exhibition coherence and visitor experience} \textbf{(RQ1)}?
To investigate this, we conduct a pilot study comparing a marker-based approach and a SLAM-based markerless approach for spatially aligning virtual artworks in an MR exhibition scenario. Second, we investigate \textit{how a deployment pipeline can support MR exhibition realization} \textbf{(RQ2)}? To answer this question, utilizing the better-performing approach identified in the pilot study, we develop a pipeline for practical exhibition deployment, covering critical steps like artwork preparation, localization, anchoring, and exhibition curation. {Finally, we explore} \textit{how the proposed pipeline in a deployed large-scale MR art exhibition shapes artwork understanding and user experience} \textbf{(RQ3)}? To answer this question, we evaluate the proposed pipeline by deploying a large-scale MR exhibition, assessing its system overhead and experiential feedback from exhibition visitors. 

Our contributions are three-fold. \textbf{First}, we provide an empirical comparison of two commonly used spatial alignment methods in an MR exhibition setting, advancing understanding of how spatial alignment choice informs exhibition design and experience. \textbf{Second}, we present a practical deployment pipeline for large-scale MR exhibitions that integrates technical deployment and exhibition curation to support practical exhibition realization. \textbf{Third}, through the real-world deployment of a large-scale MR exhibition, we contribute empirical insight with the system-level and user-centered evaluation of deployed pipeline, as well as the experiential implications for exhibition practice.

\section{Related Work}

\pb{MR Systems in Exhibition Contexts.}
Prior work has explored the deployment of MR systems in museums and exhibition settings, including cultural heritage sites and artwork installations that overlay reconstructed artifacts onto physical scenes \cite{Niko2022Design, Petrelli2019Making}. These studies demonstrate the feasibility of MR exhibition experiences, but they are typically presented as single-case prototypes or installations rather than as reusable deployment approaches. Thus, they rarely document a repeatable end-to-end process that spans preparation and on-site deployment beyond a single installation context.

Subsequent studies have extended this line of work by proposing more structured workflows for exhibition realization, particularly through authoring toolsets or curator-oriented platforms that support artwork arrangement and collaborative curation \cite{digitaltwin2021,cocreatarenhancing2025}. However, these workflows are rarely evaluated through the actual deployment of an MR exhibition, and they are seldom examined from multiple perspectives, such as system-level performance and user experience \cite{STICHELBAUT2021344, volumetricvideo2021}.

Overall, existing work either does not propose a complete pipeline for realizing MR exhibitions or lacks a sufficiently comprehensive evaluation. Our work extends this line of research by presenting a pipeline for the deployment of MR exhibitions together with a multi-perspective evaluation through a large-scale implementation in a real-world setting.

\pb{Spatial Alignment Methods for MR Exhibitions.}
Regarding spatial alignment mechanisms, existing approaches can generally be categorized as either marker-based or markerless, with SLAM-based methods being among the most commonly adopted markerless approaches \cite{reviewmixed2020}. On the one hand, marker-based methods are often used because they are easy to deploy in prototypes and require lower computational cost \cite{visualmarkerbasedlocalization2024}. However, they require visual markers (e.g., QR codes) to be placed in the physical environment, which can be visually obtrusive \cite{chicmarkerfashionably2024}. Moreover, occlusion by viewers or changes in viewing angle can reduce detection quality and even disrupt experiential continuity~\cite{augmentedreality2024}.


On the other hand, SLAM-based methods remove the need for visible physical markers by continuously mapping and localizing within the environment in real time \cite{reviewmixed2020}. Through simultaneous localization and mapping, the system estimates the device's position and orientation relative to surrounding features (e.g., corners, edges, or texture points), enabling virtual content to remain persistently anchored without relying on predefined visual markers \cite{sheng2024review}. This supports more natural and situated interaction with digital exhibits in real spaces. However, such approaches typically impose higher computational demands and are more sensitive to environmental conditions such as lighting variability \cite{reviewresearch2025}. They also suffer from drift and require re-localization in challenging environments \cite{10.3389/frobt.2022.801886, comprehensivesurvey2022}.

Although these trade-offs have been widely discussed, it remains unclear how the two spatial alignment methods affect exhibition-specific experience, particularly in terms of whether virtual content remains convincingly aligned with the physical space and whether the viewing experience remains uninterrupted as visitors move. To address this gap, we present a comparative analysis of these two methods in a simulated MR exhibition setting. This comparison also serves as the basis for the technical design of the real-world MR exhibition deployment and its subsequent evaluation.



\section{Pilot Study: Spatial Alignment Method Selection}\label{Stage_1}
To inform the design of the formal MR exhibition pipeline, a pilot user study was first conducted to determine the optimal alignment method for deployment. This preliminary evaluation compared marker-based and SLAM-based methods within a prototyped exhibition environment. This study aimed to evaluate the two methods from: \textit{spatial consistency}, \textit{experience continuity}, and the \textit{operational suitability for exhibition contexts}.

\subsection{Experiment Design and Process}
\pb{Participants.}
\label{stage_1_setting}
A total of 13 participants ($N=13$; 9 females and 4 males, denoted as P$_{p}$), aged 18–35, were recruited from the university community via a public call for volunteers. Seven participants had backgrounds in design or digital arts, while six were from engineering or computer science.

Regarding technical proficiency, the majority of the sample ($n=10$) reported limited to moderate familiarity with MR hardware, while three identified as highly experienced users. Similarly, nine participants had minimal prior exposure to immersive digital art exhibitions, whereas four considered themselves frequent visitors or experts in the medium. All experimental procedures were conducted in accordance with institutional guidelines and were formally approved by the university’s Research Ethics Committee.

\pb{Experiment settings.}
We repurposed an indoor student activity space into an MR art exhibition venue, covering an area of approximately 1,800\,m$^2$.
The exhibition featured six artworks reconstructed from open-source 3D digital assets. They were strategically arranged to simulate representative MR exhibit configurations, including assets \textit{floor-mounted}, \textit{suspended in mid-air}, and \textit{wall-aligned}.

The two aligning methods were implemented in a Unity application respectively on PICO 4 Ultra Enterprise MR headsets \cite{linowes2017augmented}.\footnote{Device specifications: \url{https://www.picoxr.com/global/products/pico4-ultra/specs}} The digital content and spatial layouts remained identical across both conditions. Figure~\ref{fig:pipeline_offline} and~\ref{fig:pipeline_online} illustrates the architectures of deployed SLAM-based method. For the marker-based method, we employed a commonly used architecture in prior MR studies~\cite{dressler2018mediating,visualmarkerbasedlocalization2024,chicmarkerfashionably2024}.

\pb{Experimental protocol.}
All participants experienced the MR exhibition under both aligning methods. To mitigate order effects, the presentation sequence was counterbalanced across the cohort ~\cite{field2024discovering}.
Participants were instructed to navigate the exhibition space and engage with the digital assets freely, observing the 3D content from multiple perspectives to simulate an authentic gallery visit.

Following each trial, participants provided quantitative feedback using an adapted Igroup Presence Questionnaire (IPQ)~\cite{schubert2001experience} and a User Experience Questionnaire (UEQ)~\cite{laugwitz2008construction}.  
After completing both trials, participants completed a comparative survey evaluating the two methods in terms of operational suitability~\cite{wanzer2020experiencing}. Both questionnaires used a five-point Likert scale, with a score of 5 indicating the strongest agreement.

The study of each user concluded with a semi-structured individual interview focused on the perceived visual stability, smoothness, and experiential impact of each method. The interview was structured around three questions (detailed in Supplementary Materials) and all responses were recorded and subsequently transcribed for qualitative analysis.

\pb{Data analysis methods.}
We performed a quantitative comparative analysis of the questionnaire responses to evaluate the suitability of each method. Given the small sample size ($N=13$), we bypassed formal normality testing (e.g., Shapiro-Wilk), as such tests often lack the statistical power to reliably detect deviations in small cohorts. Consequently, we employed the Wilcoxon signed-rank test, a non-parametric approach well-suited for ordinal Likert-scale data and paired-sample comparisons without assuming a normal distribution. In addition, all questionnaires possess a Cronbach's $\alpha>0.7$, showing acceptable reliability (see Table~\ref{tab:pilot_ipq_ueq} and~\ref{tab:pilot_intra_compare}).


As for interview responses, we conducted a thematic analysis of interview transcripts following Braun and Clarke’s six-phase framework~\cite{braun2006using}. After a period of becoming familiar with the data through multiple readings of the responses, one author generated 24 initial open codes covering users' technical and experiential attitudes. These were refined by a second author to ensure inter-coder reliability. Through a series of iterative consensus-building joint sessions, we synthesized these into seven themes across three dimensions of the MR exhibition experience.





\begin{table*}[t]
\centering
\resizebox{.9\linewidth}{!}{%
\begin{tabular}{@{}|l|l|l|cc|cc|l|@{}}
\toprule
\multirow{2}{*}{\textbf{Aspect to evaluate}} & \multirow{2}{*}{} & \multirow{2}{*}{\textbf{Question}} & \multicolumn{2}{c|}{\textbf{Marker}} & \multicolumn{2}{c|}{\textbf{SLAM}} & \multirow{2}{*}{\textbf{\textit{W}}} \\
\cmidrule(lr){4-5}\cmidrule(lr){6-7}
& & & \textbf{Mdn/IQR} & \bm{$\alpha$} & \textbf{Mdn/IQR} & \bm{$\alpha$} & \\
\midrule
\multirow{4}{*}{\shortstack[l]{Spatial Presence\\Immersive Experience\\(IPQ-adapted)}} & Q$_{p}$1 & I felt present in the mixed (physical + virtual) space. & 3 / 1 & \multirow{4}{*}{0.81} & 5 / 1 & \multirow{4}{*}{0.84} & 3.0** \\
& Q$_{p}$2 & I had the feeling of really being there in the mixed environment. & 3 / 1 & & 5 / 1 & & 0.0** \\
& Q$_{p}$3 & The virtual assets felt naturally integrated into the physical scenes. & 3 / 1 & & 4 / 0 & & 0.0** \\
& Q$_{p}$4 & My exploration in the space felt natural and smooth. & 3 / 1 & & 4 / 1 & & 0.0*** \\
\midrule
\multirow{8}{*}{\shortstack[l]{System Usage Experience\\(UEQ-S)}}& Q$_{p}$5 & The system was easy to use. & 4 / 1 & \multirow{8}{*}{0.90} & 5 / 1 & \multirow{8}{*}{0.92} & 4.5** \\
& Q$_{p}$6 & The system is efficient. & 3 / 1 & & 5 / 1 & & 5.0** \\
& Q$_{p}$7 & The system is exciting to use. & 3 / 0 & & 4 / 1 & & 0.0** \\
& Q$_{p}$8 & The system is fun to use. & 4 / 2 & & 4 / 1 & & 3.0* \\
& Q$_{p}$9 & The system is pleasant to use. & 3 / 2 & & 4 / 1 & & 0.0** \\
& Q$_{p}$10 & The system has a pleasant appearance. & 3 / 1 & & 4 / 1 & & 0.0** \\
& Q$_{p}$11 & The system is innovative. & 3 / 2 & & 5 / 1 & & 2.5* \\
& Q$_{p}$12 & The system is original. & 4 / 1 & & 4 / 1 & & 13.5 \\ \bottomrule
\end{tabular}%
}
\caption{A summary of the IPQ and UEQ results in pilot study. ``Mdn/IQR'' denotes the median value and interquartile range of participants' scores to corresponding questions. \bm{$\alpha$} denotes the Cronbach's Alpha mapped to corresponding aspect to evaluate. The statistical significance is reported by the Wilcoxon signed-rank test. *: $p<0.05$; **: $p<0.01$; ***: $p<0.001$.}
\label{tab:pilot_ipq_ueq}
\end{table*}

\begin{table}[t]
\centering
\resizebox{\linewidth}{!}{%
\begin{tabular}{@{}|c|l|L{25em}|c|c|@{}}
\toprule
 &  & \textbf{Question} & \textbf{Mdn/IQR} & \bm{$\alpha$} \\ \midrule
\multirow{12}{*}{\rotatebox{90}{Intra-method comparison}} & Q$_{p}$13 & As a visitor, I prefer using SLAM-based method if in a practical MR exhibition. & 5 / 0 & \multirow{12}{*}{0.794} \\
 & Q$_{p}$14 & SLAM-based method is easier to experience. & 5 / 1 &  \\ 
 & Q$_{p}$15 & SLAM-based method facilitates my understanding of the artwork. & 4 / 2 &  \\
 & Q$_{p}$16 & SLAM-based method enhances the artwork expression. & 5 / 1 &  \\
 & Q$_{p}$17 & SLAM-based method does not distract me from the artwork. & 4 / 2 &  \\
 & Q$_{p}$18 & SLAM-based method feels comfortable for long-time exhibition viewing. & 4 / 1 &  \\
 \cmidrule(lr){2-4}
 & Q$_{p}$19 & For an organizer, SLAM-based method is more suitable for exhibition organization. & 5 / 0 &  \\
 & Q$_{p}$20 & SLAM-based method can function reliably in a public exhibition setting. & 4 / 2 &  \\
 \bottomrule
\end{tabular}%
}
\caption{Questionnaire results for intra-method comparison.}
\label{tab:pilot_intra_compare}
\end{table}

\subsection{Experiment Results}\label{Pilot_Result}

\pb{Questionnaire results.} Table~\ref{tab:pilot_ipq_ueq} summarizes the performance comparison between marker-based and SLAM-based methods reported by IPQ and UEQ-S questionnaires. Overall, we find participants express agreement on that SLAM-based method can provide significantly \textit{better} immersion and system usage experience in a MR exhibition setting.

According to IPQ results, participants reported with a significantly higher level of agreement on that SLAM-based method could foster the feeling of presence (Q$_{p}$1: median score 5 vs. 3, for SLAM vs. marker, $W=3.0$, $p<0.01$) and immersion (Q$_{p}$2: 5 vs. 3, $W=0.0$, $p<0.01$) than the case with marker-based method. Moreover, participants also agreed more on that SLAM-based method could facilitated a natural integration of virtual assets and physical scenes (Q$_{p}$3: 4 vs. 3, $W=0.0$, $p<0.01$), which supported a natural and smooth visitor experience (Q$_{p}$4: 4 vs. 3, $W=0.0$, $p<0.001$).

As for UEQ-S results, participants showed significantly higher agreement level on the ease (Q$_{p}$5: 5 vs. 4, $W=4.5$, $p<0.01$) and efficiency (Q$_{p}$6: 5 vs. 3, $W=5.0$, $p<0.01$) of using SLAM-based method than the marker-based one. Meanwhile, participants also agreed that they were more like to feel excitement (Q$_{p}$7: 4 vs. 3, $W=0.0$, $p<0.01$) and pleasure (Q$_{p}$9: 4 vs. 3, $W=0.0$, $p<0.01$) when using SLAM-based method. This can be related to their appreciation towards corresponding system appearance (Q$_{p}$10: 4 vs. 3, $W=0.0$, $p<0.01$) and SLAM-based method is considered as more innovative (Q$_{p}$11: 5 vs. 3, $W=2.5$, $p<0.05$).

Additionally, Table~\ref{tab:pilot_intra_compare} presents the results of intra-method comparison questionnaire to investigate which method users prefer when acting as a visitor and a organizer. On the one hand, form the visitor's perspective, users demonstrated a notable leaning towards SLAM-based method (Q$_{p}$13: median score 5), as well as a strong agreement on its ease of usage (Q$_{p}$14: 5). Particularly, users highlighted the effect of SLAM-based method in facilitating visitors' understanding (Q$_{p}$15: 4) towards artwork by enhancing its art expression (Q$_{p}$16: 5). Additionally, the SLAM-based method also introduced less distraction (Q$_{p}$17: 4) and discomfort (Q$_{p}$18: 4) during a long-time exhibition viewing. On the other hand, form the organizer's aspect, the SLAM-based method was considered as more suitable (Q$_{p}$19: 5) and reliable (Q$_{p}$20: 4) for exhibition organization.


\pb{Interview results.}
Our qualitative analysis on participants' interview responses revealed three aspects related to their evaluation to the two aligning methods.

\spb{Aspect 1: Presence inconsistency.} In the interview, 9 participants (69.23\%) highlighted the presence inconsistency issue that often occurred when they viewing exhibits.
\one A total of 5 participants (38.46\%) reported significant issues with the \textbf{initial placement and orientation of the virtual content}. This problem was predominantly observed in the marker-based method (4 participants, 30.77\%), where the offsets led to spatial disconnects. Participants noted that exhibits often failed to align with the physical environment, with P$_{p}$10 remarking, "\textit{The artwork was far from the marker... the offset was very severe.}'' Others experienced depth or placement errors, such as the artwork appearing to be "\textit{stuck in the wall}" (P$_{p}$08) or \textit{"floating"} rather than being pressed tightly against the wall'' (P$_{p}$01). In one extreme case, an exhibit was found to have "drifted outside the door" (P$_{p}$10). In contrast, only 1 participant (7.69\%) reported a similar issue with the SLAM-based method, describing a sudden scaling error where the exhibit appeared to "\textit{shrink... as if moving further away}" (P$_{p}$02).
\two Beyond static placement, \textbf{the stability of the exhibits during participant movement} was a recurring concern. Participants frequently encountered ``jumping" or ``shaking" exhibits that disrupted the immersion. For instance, P$_{p}$04 noted that upon approaching an exhibit, ``\textit{it jumped to a different position.}'' High-frequency jitter was also reported by P$_{p}$06, who observed the content ``\textit{shaking back and forth... two or three times a second}'' before stabilizing. These stability issues were also present in the SLAM-based method (2 participants, 15.38\%). Participant reported lateral drift and depth inconsistency during active exploration: ``\textit{When turning my head, it would drift and this made the inconsistency more noticeable.}'' (P$_{p}$09).

\spb{Aspect 2: Experience interruption.}
Beyond spatial inconsistencies, 5 participants (38.46\%) reported significant \textbf{experience interruptions specifically linked to the marker-based method}. These participants felt that the necessity of scanning physical markers shifted their focus away from the artistic content, with P$_{p}$01 noting that they were often ``\textit{looking for the marker rather than viewing the artworks}'' and felt compelled to ``\textit{walk to the spot}'' just to trigger an exhibit. This reliance on external triggers introduced both visual and technical friction; P$_{p}$10 described the markers as an ``\textit{interference element}'' unrelated to the exhibit, while P$_{p}$11 expressed frustration at having to ``\textit{look all around}'' to locate them. Furthermore, the sensitivity of the alignment caused the experience to feel fragmented, as markers would frequently go ``\textit{in and out}'' of the field of view, which P$_{p}$07 identified as a primary factor interrupting their immersion. This was compounded by latent loading times after successful scans, leading to a ``\textit{scanning-heavy}'' interaction that many participants found detrimental to the overall experience.

\spb{Aspect 3: Goodness for implementation.} Finally, participants' feedback also highlighted the suitability for implementation, with a strong preference for the coherence offered by the SLAM-based approach. \one 6 participants (46.15\%) specifically praised the SLAM method for \textbf{maintaining exhibition coherence}, noting that the absence of a ``one-by-one'' scanning requirement eliminated the fragmented nature of the marker-based experience. As P$_{p}$02 observed, the lack of forced pauses allowed scenes to ``\textit{follow one another}'' naturally, creating a ``\textit{sense of wholeness}'' that P$_{p}$05 felt led to a superior understanding of the exhibition. This spatial fluidity allowed exhibits to appear naturally distributed, enabling users like P$_{p}$11 to approach them smoothly without the distraction of searching for codes. \two Consequently, 7 users (53.84\%) found the SLAM-based method to be \textbf{more complete and immersive}; P$_{p}$08 contrasted this "\textit{museum-like experience}" and its ``\textit{stronger sense of place}'' against the marker-based method, which felt more like "\textit{scanning for a discount or a deal}.'' \three However, the marker-based approach retained some \textbf{perceived advantages for implementation}, with 3 users (23.08\%) noting its efficiency in terms of setup, as it required ``\textit{no prior scene scanning and data processing}'' (P$_{p}$06) before the experience began.

\pb{Aligning method selection.} Overall, the SLAM-based method offers a more suitable foundation for practical MR exhibition workflows. In contrast to the marker-based method, which introduced frequent experience interruptions and fragmented user engagement, SLAM-based method better supports continuous spatial alignment and minimizes disruptive interactions. This aligns with participants' consistent emphasis on exhibition coherence and a sustained sense of immersion. Given our goal of designing a workflow that preserves exhibition coherence and enhances users' immersive experience, we therefore select the SLAM-based method as the primary method.

\section{Pipeline Design for MR Exhibition}

Based on the pilot study results, we selected the SLAM-based method for the formal MR exhibition deployment. In this section, we present the deployment pipeline designed for the exhibition. We also describe how it was used to realize a large-scale MR exhibition in a real-world setting. The goal was to develop a technical and curatorial approach that could support a coherent exhibition experience.

\begin{figure}[t]
   \centering
   \includegraphics[width=\linewidth]{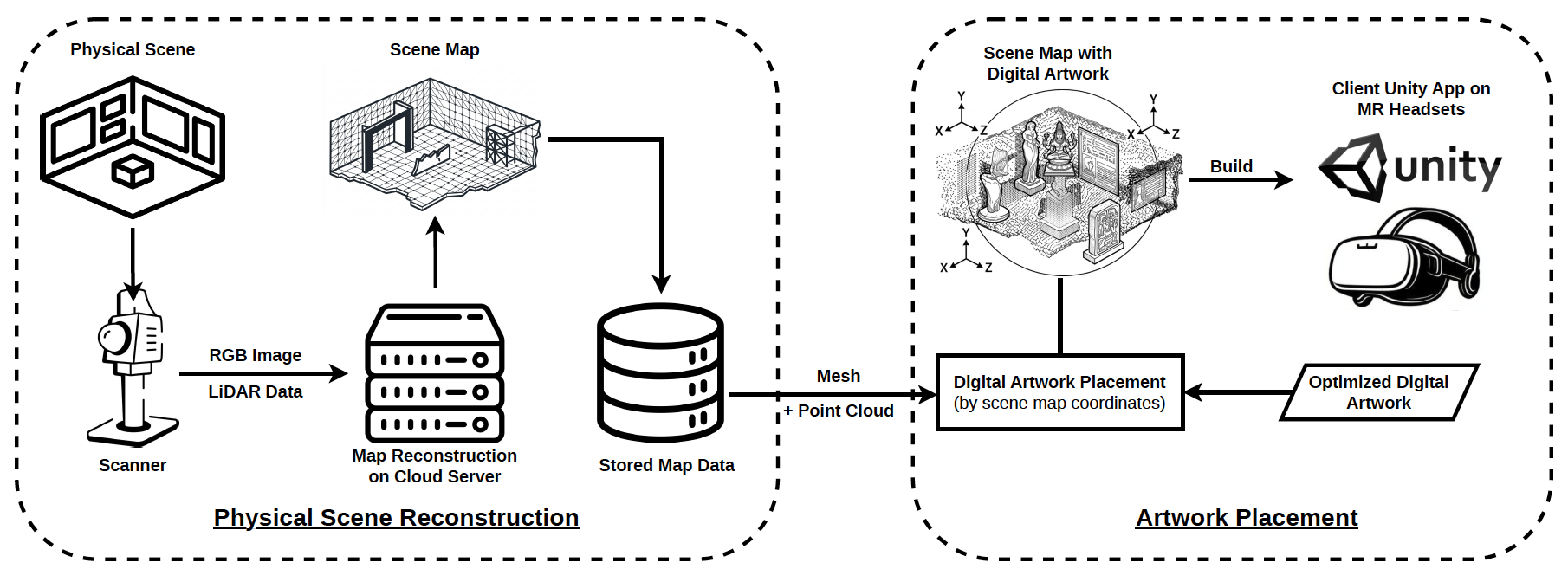}
   \caption{Pipeline steps of physical scene construction and artwork placement (steps that incorporate with exhibition curation and design.)}
   \Description{}
   \label{fig:pipeline_offline}
 \end{figure}

\pb{\uline{Physical Scene Reconstruction} and \uline{Artwork Placement}.}
The pipeline’s technical deployment began with physical scene reconstruction and artwork placement, which served as preparatory steps for the subsequent MR exhibition experience. As shown in Figure \ref{fig:pipeline_offline}, the physical site was first scanned, and the captured RGB images and LiDAR data were transmitted to a server to reconstruct the scene map. In parallel, the digital artworks were prepared and optimized in Blender or Unity through format conversion, shader configuration, compression, and performance tuning \cite{blender_manual, unity_manual}. Afterwards, the reconstructed mesh and point cloud of the scene-map data were used as spatial references for placing the artworks according to scene-map coordinates, which also encoded the geometric and spatial structure of the site. Finally, the scene map and the placed artworks were built into the client Unity application on the MR headset.

\begin{figure}[t]
   \centering
   \includegraphics[width=\linewidth]{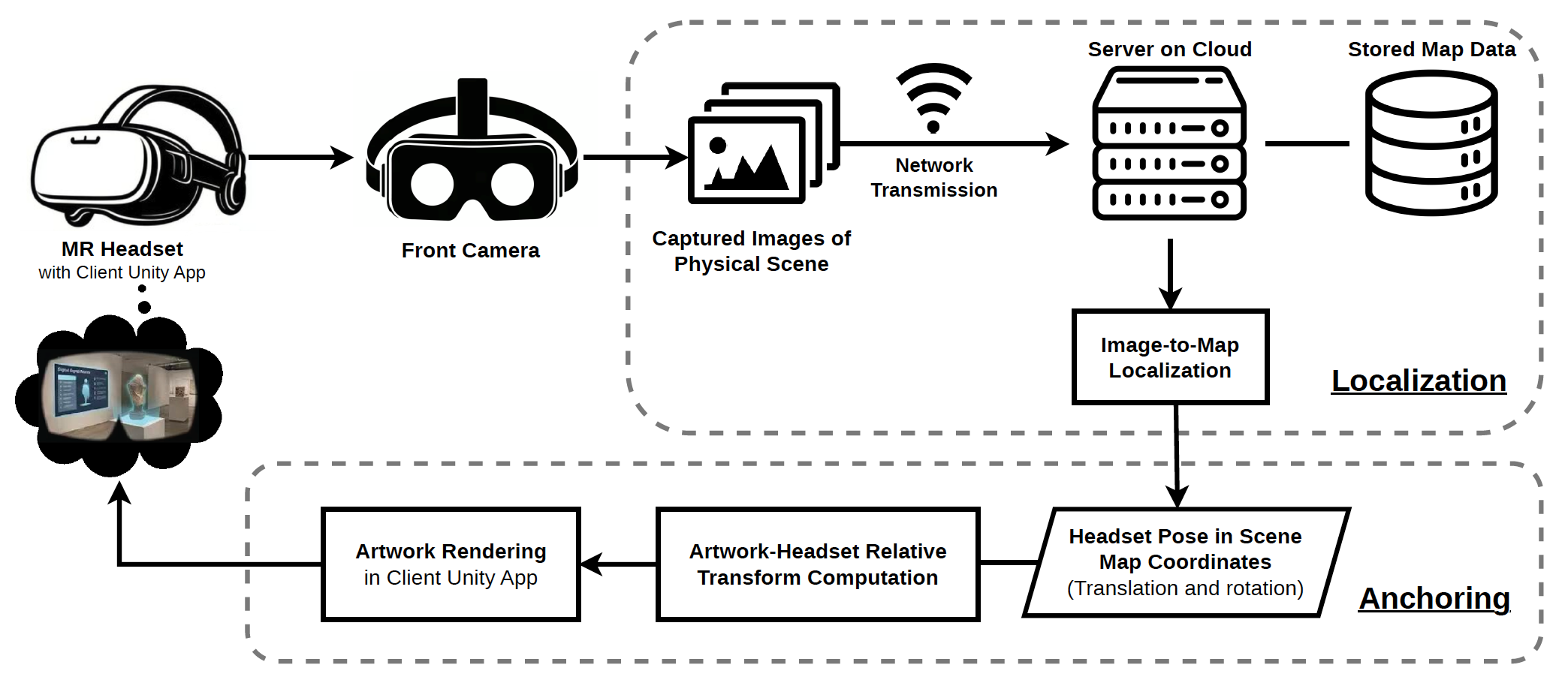}
   \caption{Pipeline steps of headset pose localization and artwork anchoring (steps that incorporate with viewer experience.)}
   \Description{\yawei{Placeholder, replace later}}
   \label{fig:pipeline_online}
 \end{figure}

\pb{\uline{Localization} and \uline{Anchoring}.}
Headset pose localization and artwork anchoring were the runtime stages of the technical deployment, supporting exhibition viewing by enabling digital artworks to be stably overlaid onto the physical environment in real time. As illustrated in Figure \ref{fig:pipeline_online}, when users visited the exhibition, the MR headset cameras first captured images of the physical scene, which were then transmitted to the server. On the server side, these images were used to localize the headset pose within the scene-map coordinates. Specifically, the system estimated the pose of the headset camera, including its translation and rotation, corresponding to the headset’s position and orientation. Accordingly, the server computed the relative spatial relationship between the headset and the placed artworks through relative transformation. Subsequently, the artworks were rendered in the client Unity application and displayed through the headset at locations corresponding to their placement in the scene map.


\pb{\uline{Exhibition Curation}.}
Exhibition curation formed a parallel but conceptually distinct part of the deployment process in our pipeline. It proceeded alongside the technical deployment and provided curatorial decisions to guide the placement of digital artworks in the reconstructed physical scene. Following established curatorial methodologies \cite{o2016culture,bogle2013museum}, this process involved three main steps: thematic formation, spatial design with narrative alignment, and artwork--place matching. 

First, the overall theme and sub-themes were defined based on the curatorial intentions and the conceptual content of the 31 digital artworks selected through an open call. Based on these themes and the meanings conveyed by the artworks, we developed thematic narratives for the exhibition. In parallel, we selected the exhibition area and divided it into four connected zones. Each zone was aligned with a corresponding narrative sub-theme. Artwork clusters under each sub-theme were then assigned to the different zones.

\begin{figure}[t]
   \centering
   \includegraphics[width=\linewidth]{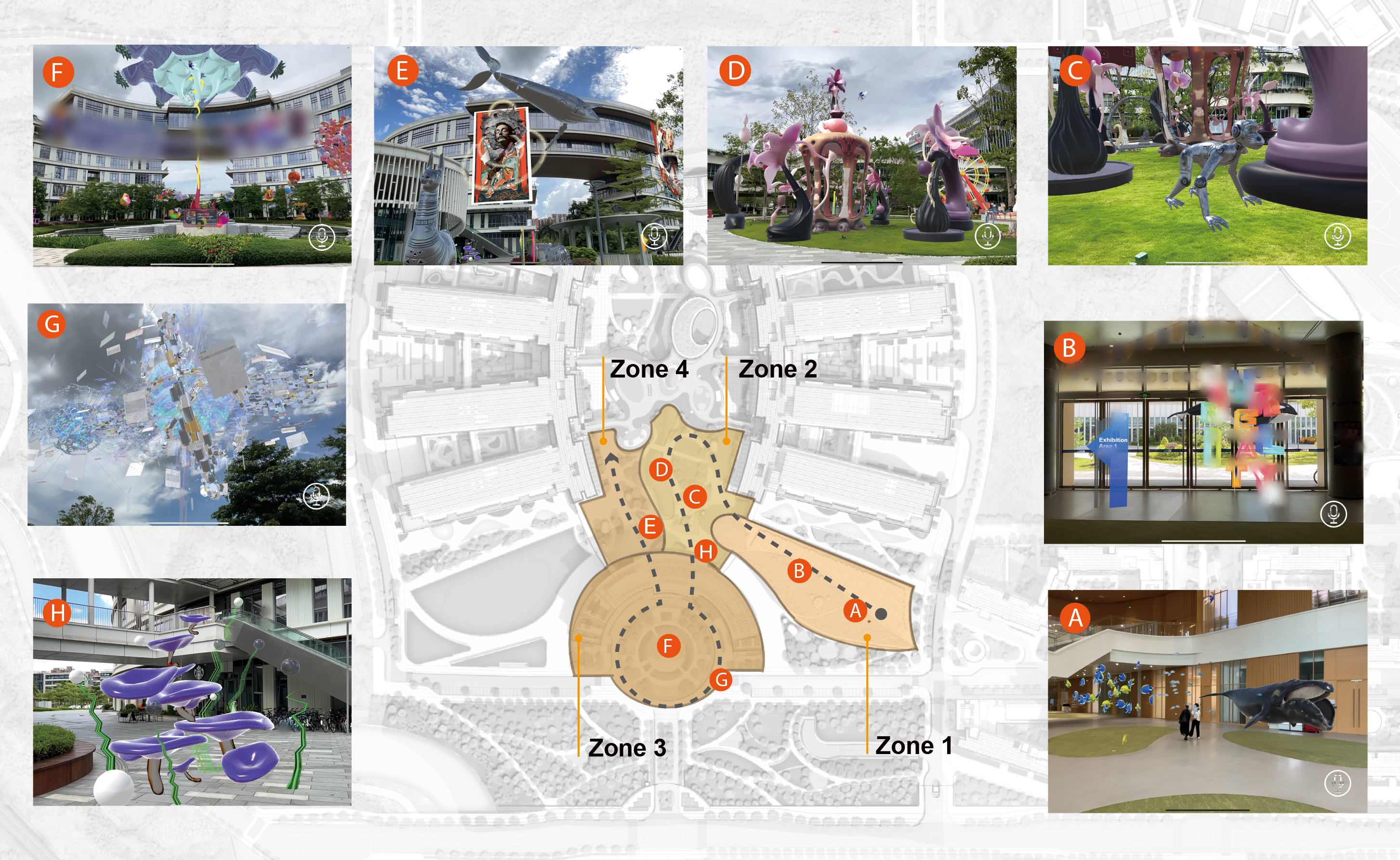}
   \caption{The layout of exhibition zones and representative deployed artworks within the MR exhibition.}
   \Description{}
   \label{fig:pipeline_curation}
 \end{figure}

We then conducted artwork—physical matching. Each artwork was assigned to a specific location, or digital artwork--physical interface. This assignment was determined by the correspondence between the artwork’s geometric or narrative characteristics and those of the physical site. For example, poster-like planar artworks were matched with building facades of similar planar form (thumbnail E in Figure \ref{fig:pipeline_curation}), while plant-like or animal-like artworks were placed in lawn areas (thumbnail C or D in Figure \ref{fig:pipeline_curation}).

Finally, we added supportive design elements as supplementary features to enrich the exhibition experience. These included multimedia effects such as directional sound cues, navigational signs, and clickable information labels. For the final deployment, the exhibition covered approximately 26,000 m$^2$. It was organized into four exhibition zones connected by a designed viewing flow. The organizational details, together with representative artwork images, are shown in Figure \ref{fig:pipeline_curation}.



\section{Pipeline Evaluation}
To evaluate the performance of the practical MR exhibition pipeline, we conducted a formal study to assess: \textit{system overhead} and \textit{user experiential feedback}.

\subsection{Experiment Design and Process}

\pb{Participants}
A total of 30 participants ($N=30$; 18 females, 12 males, denoted as P$_{f}$), aged 18 to 35, were recruited from the university community via a public call for volunteers.
The majority of the sample ($n=21$) reported limited to moderate familiarity with MR hardware, while nine identified as highly experienced users. 
In addition, 20 participants had moderate or higher knowledge level on art and exhibition, whereas ten considered themselves unfamiliar with this venue. 
All experimental procedures were formally approved by the university’s Research Ethics Committee.

\pb{Experiment settings.}
Unlike the pilot study, which simulated exhibit arrangement, the formal study was conducted in a real MR art exhibition setting. Figure~\ref{fig:pipeline_curation} shows a plan view of the exhibition zones, visit routes, and artwork distribution. Participants were organized into tour groups of up to five. Each participant completed an exhibition visit of approximately 20 minutes using a client Unity application deployed on PICO 4 Ultra Enterprise MR headsets.

\pb{Experimental protocol.}
Participants first engaged in the MR exhibition by following the route, accompanied by a tour guide to ensure navigational safety. To evaluate system performance and overhead, the Unity client application logged four key telemetry metrics: \textit{Frames Per Second (FPS)} \cite{wang2023effect}, \textit{Frame Time} \cite{visser2015dynamics}, \textit{CPU Time per Frame} \cite{qasaimeh2019comparing}, and \textit{Memory Usage}. Data collection commenced at a sampling rate of 1 Hz within the 20 minutes viewing, immediately following the successful completion of asset localization.

Upon completion of the exhibition, participants were administered three standardized questionnaires to evaluate the system’s performance from a user-centric perspective. To assess perceived cognitive workload, we utilized an adapted NASA Task Load Index (NASA-TLX)~\cite{advancingnasatlx2024}. User experience and system usability were quantified using the UEQ~\cite{laugwitz2008construction, userexperience2021} and the System Usability Scale (SUS)~\cite{Lewis2018system}, respectively. All three questionnaires were implemented using five-point Likert scales, where a score of 1 indicated the lowest performance (e.g., very high workload, very poor usability) and 5 indicated optimal performance.

Afterwards, participants were invited to take an optional 20-minute semi-structured interview containing eight questions (detailed in Supplementary Materials).
These questions were designed to explore the perceived integration of digital content within the physical environment and the impact of system stability on user immersion \cite{cummings2016immersive, al2022framework}. Specifically, we sought to evaluate whether the MR experience was perceived as a cohesive art exhibition and other experiential insights.
Finally, 18 users attended the interview and all responses were recorded and transcribed for qualitative analysis.

\pb{Data analysis methods.}
We employ similar methods as in pilot study (Section~\ref{stage_1_setting}) to analysis both questionnaire and interview responses collected in the formal study. All questionnaires possess a Cronbach's $\alpha>0.7$, showing acceptable reliability. For the qualitative analysis, we first generated 205 initial codes and finally synthesized these into four themes across two dimensions.

\subsection{Results}

\begin{figure*}[t]
    \centering
    \begin{subfigure}[b]{.24\linewidth}
        \centering
        \includegraphics[width=\linewidth]{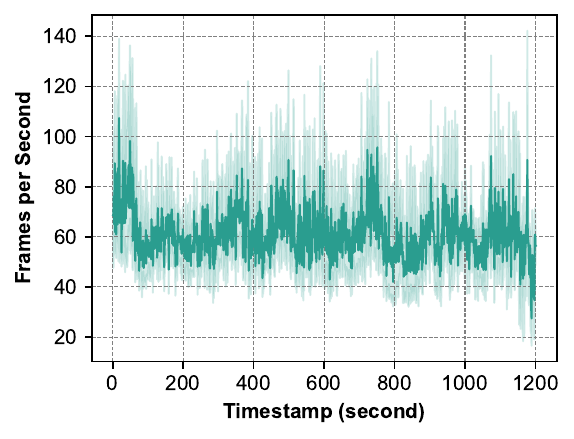} 
    \end{subfigure}
    \begin{subfigure}[b]{.24\linewidth}
        \centering
        \includegraphics[width=\linewidth]{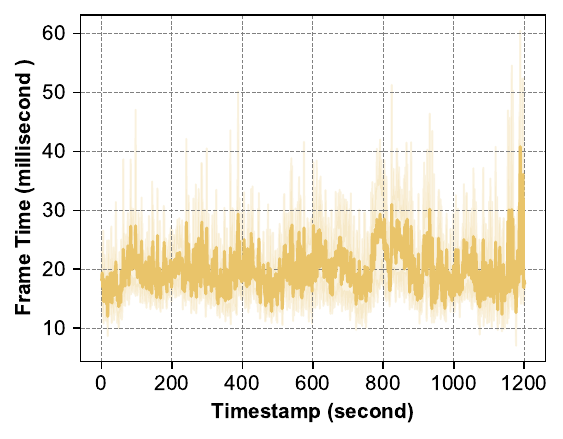}  
    \end{subfigure}
    \begin{subfigure}[b]{.24\linewidth}
        \centering
        \includegraphics[width=\linewidth]{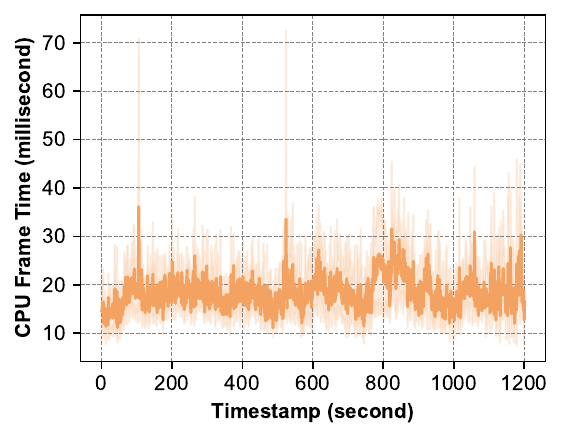}
    \end{subfigure}
    \begin{subfigure}[b]{.24\linewidth}
        \centering
        \includegraphics[width=\linewidth]{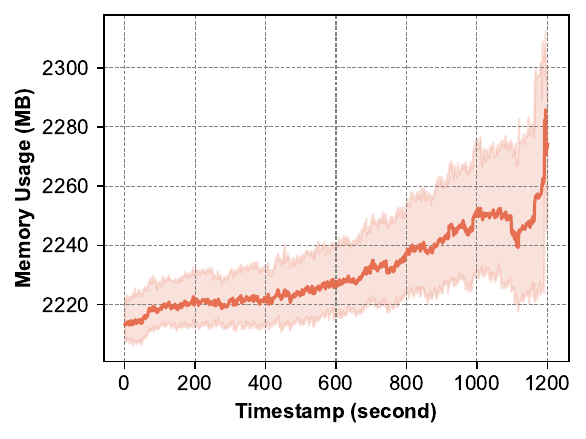}
    \end{subfigure}
    \caption{The temporal distribution of metrics to evaluate the system overhead of the proposed pipeline.}
    \label{fig:system_eval}
\end{figure*}

\pb{System overhead.}
Figure~\ref{fig:system_eval} demonstrates the temporal distribution of the four evaluative metrics throughout the 20-minute viewing session. In general, our results demonstrates that the MR exhibition deployed based on our pipeline can maintain system stability and responsiveness. 

On a Pico 4 Ultra headset, the exhibition achieved a frame rate between 60 and 80 FPS, generally above the threshold required for a fluid and comfortable user experience in spatial computing~\cite{pico_pxrc}. This was further supported by the consistently low frame and CPU processing times, which remained stable around 20 ms with minimal jitter. This low computational overhead ensures a highly responsive system capable of near-instantaneous updates as users move throughout the exhibition space. Although memory usage shows a gradual linear increase from approximately 2215 MB to 2280 MB, the total footprint remains a small fraction of the device's 12GB capacity. Considering the exhibition simultaneously managed the virtual assets and spatial anchors for 21 distinct artworks, this suggests that the system can handle prolonged exhibition cycles without immediate risk of thermal throttling or performance degradation. Overall, these results imply a robust hardware-software synergy that allows the deployed exhibition to remain unobtrusive, effectively supporting the immersive and artistic goals.

\begin{table}[t]
\centering
\resizebox{\linewidth}{!}{%
\begin{tabular}{@{}|c|l|L{25em}|c|c|@{}}
\toprule
 &  & \textbf{Question} & \textbf{Mdn/IQR} & \bm{$\alpha$} \\ \midrule
\multirow{6}{*}{\rotatebox{90}{NASA-TLX}} & Q$_{f}$1* & How much mental activity was required? & 4 /2  & \multirow{6}{*}{0.854} \\
 & Q$_{f}$2* & How much physical activity was required? & 4 /1  &  \\ 
 & Q$_{f}$3* & How much time pressure did you feel in the task? & 5 /1  &  \\
 & Q$_{f}$4* & How irritated, stressed, and annoyed did you feel in the task? & 4.5 /1  &  \\
 & Q$_{f}$5* & How hard did you have to work to accomplish your level of performance? & 4 /0.75  &  \\
 \midrule
 \multirow{6}{*}{\rotatebox{90}{UEQ}}& Q$_{f}$6 & How enjoyable the MR exhibition is? & 4 / 1 & \multirow{6}{*}{0.747} \\ 
 & Q$_{f}$7 & How clear the MR exhibition is? & 4 / 2 &  \\
 & Q$_{f}$8 & How attractive the MR exhibition is? & 4 / 1.75 &  \\
 & Q$_{f}$9 & How easy the exhibition application is to use? & 4 / 1 &  \\
 & Q$_{f}$10 & How exciting the exhibition is? & 4 / 2 &  \\
 & Q$_{f}$11 & How inventive the MR exhibition is? & 4 / 1 &  \\
 \midrule
 \multirow{13}{*}{\rotatebox{90}{SUS}}& Q$_{f}$12 & I think that I would like to use this system frequently. & 4 / 1 & \multirow{13}{*}{{0.805}} \\
 & Q$_{f}$13* & I find the system unnecessarily complex. & 4 / 1 &  \\
 & Q$_{f}$14 & I think the system is easy to use. & 4 / 0.75 &  \\
 & Q$_{f}$15* & I think that I would need the support of a technical person to be able to use this system. & 4 / 1 &  \\
 & Q$_{f}$16 & I find the various functions in this system are well integrated. & 4 / 0 &  \\
 & Q$_{f}$17* & I think there is too much inconsistency in this system. & 4 / 2 &  \\
 & Q$_{f}$18 & I will imagine that most people will learn to use this system very quickly. & 4 / 2 &  \\
 & Q$_{f}$19* & I find the system very cumbersome (awkward) to use. & 4 / 2 &  \\
 & Q$_{f}$20 & I feel very confident using the system. & 4 / 1.75 &  \\
 & Q$_{f}$21* & I need to learn a lot of things before I can get going with this system. & 4 / 1 &  \\
 \bottomrule
\end{tabular}%
}
\caption{Questionnaire results for formal user study. *: the scores of negative items were reversed for the ease of evaluation, where 5 always indicated optimal performance.}
\label{tab:formal_questionnaire}
\end{table}

\pb{Questionnaire results.} Table~\ref{tab:formal_questionnaire} summarizes the results of the formal user study across three standardized dimensions: workload (NASA-TLX), user experience (UEQ), and system usability (SUS). 

First, regarding the workload and cognitive demand (NASA-TLX), our pipeline performed consistently well, with all reversed scores of 4 or higher. Notably, users reported a low level of time pressure ($Q_{f}3$: median score 5) and minimal irritation or stress ($Q_{f}4$: 4.5) while interacting with the MR exhibition. These results imply that the SLAM-based interaction is intuitive enough to prevent cognitive overload, even in a complex exhibition environment.

Second, the user experience (UEQ) scores reflect a balanced and positive reception. Users found the exhibition to be enjoyable ($Q_{f}6$: 4), inventive ($Q_{f}11$: 4), and easy to use ($Q_{f}9$: 4). The consistency across these metrics suggests that our pipeline successfully bridges the gap between technical utility and aesthetic engagement, maintaining a high level of excitement and attractiveness alongside functional clarity.

Finally, the system usability (SUS) results further validate the robustness of our pipeline. Users expressed a strong desire for frequent use ($Q_{f}12$: 4) and highly valued the integration of various functions ($Q_{f}16$: 4). The high reversed scores for complexity ($Q_{f}13$: 4) and need for technical support ($Q_{f}15$: 4) indicate that our pipeline is perceived as self-explanatory and accessible to non-expert visitors. Furthermore, a median score of 4 for $Q_{f}20$ suggests that the SLAM-based approach instills a high degree of user confidence, which is critical for autonomous navigation in public MR spaces.

\pb{Interview results.}
Our qualitative analysis of the 18 participant interviews (P$_{f}$01--P$_{f}$18) revealed following primary aspects regarding the perception of the MR exhibition deployed by our pipeline.


\spb{Aspect 1: Artwork-space integration.} Beyond mere placement, the SLAM-based method was described as a tool for extending the physical space, creating ``\textit{dynamic connections}'' between digital artworks and physical space (P$_{f}$02). This effectiveness often relied on how artworks ``responded'' to specific site characteristics, such as an exhibit situated at a corridor corner that ``\textit{echoed the architectural transition of the building}'' (P$_{f}$05). Participants also emphasized that the successful integration was not accidental but attributed to the deep curation of exhibition organization, particularly on "\textit{the scale, form, and content logic in relation to the environment}'' (P$_{f}$11).

\spb{Aspect 2: Content-location synergy.} Participants' responses also suggest that, in the MR exhibition, memory towards artworks is not a choice between content and location but rather a product of their synergy. Participants reported a high capacity for recalling artwork content, with P$_{f}$06 noting the ability to ``\textit{completely recall}'' which specific artwork appeared at which physical landmark. This suggests that physical landmarks act as powerful mnemonic anchors for digital content. Memory was most potent when the artwork and environment achieved a ``\textit{contextual fit}'' (P$_{f}$03, P$_{f}$07). When an exhibit effectively leveraged its surroundings as context, the two elements were encoded as ``\textit{a single, inseparable unit}'' (P03, P07). This interaction was even described by P08 as a "\textit{new work}" in itself, a hybrid entity where the connection between the digital and physical becomes the primary object of memory, rather than the isolated digital asset.

\section{Discussion}
\subsection{Spatial Alignment as a Foundational Design Decision for MR Exhibitions (RQ1)}

Our comparative analysis suggests that, in MR exhibitions, method selection should be informed not only by technical alignment performance but also, more importantly, by its implications for exhibition coherence and experience. Although both methods could exhibit spatial inconsistencies, participants more often associated the marker-based method with severe initial placement and orientation offsets, repeated scanning demands, and a fragmented viewing process. These issues were not experienced as isolated technical flaws, but as disruptions to exhibition continuity. By contrast, the SLAM-based method was more often described as supporting smoother transitions between exhibits, a stronger "\textit{sense of wholeness,}" and a more immersive and "\textit{natural museum-like}" experience. Participants also more frequently regarded it as a more suitable basis for exhibition deployment because it better sustained the spatial and experiential coherence of the exhibition as a whole.


Furthermore, spatial alignment matters not simply as a technical property, but through the way it shapes exhibition coherence, continuity, and immersion. Participants described these effects in experiential terms, such as whether artworks "\textit{stayed fixed,}" whether movement introduced "\textit{drift or jitter,}" and whether the exhibition felt continuous or interrupted. This connection was also evident in the formal interviews: when content remained stable and spatially fixed, the exhibition was experienced as more continuous and immersive, with the technology receding into the background; when drift, clipping, or instability occurred, attention was redirected to the system itself, and disruptions to exhibition coherence became more salient. Taken together, these findings suggest that spatial alignment is not merely an implementation choice in MR exhibitions, but a design factor that impacts the exhibition coherence across space and viewers' sense of continuity and immersion. Therefore, for curators and designers of MR exhibitions, we highlight that spatial alignment choice should be considered holistically alongside exhibition design.

\pb{Take-home message:} Spatial alignment method selection is not merely a technical choice in MR exhibitions, but a design decision that can shape exhibition coherence and viewers' sense of continuity and immersion.


\subsection{Pipeline Design Implication for Coherent MR Exhibition Realization (RQ2)}

Coherent MR exhibition realization requires a deployment pipeline in which technical preparation and exhibition curation work together. In our pipeline, physical scene reconstruction and artwork placement established the spatial reference for deployment, headset pose localization and artwork anchoring supported real-time on-site viewing, and exhibition curation guided how artworks were organized and coherently matched to the site and narrative themes. The value of the pipeline, therefore, lay not only in enabling practical deployment of the SLAM-based method in an exhibition setting, but also in allowing technical deployment and curatorial strategy to operate as one coordinated process.

The formal study suggests that coherent MR exhibition realization depended on more than accurate runtime registration. Rather, coherence emerged when the deployment pipeline preserved the spatial and curatorial relationships established before the exhibition. Artworks were perceived as more convincing when meaningfully placed in relation to the surrounding site and encountered naturally along the intended visitor route. Runtime alignment helped preserve these relationships during on-site viewing. This coordination was supported by both technical and qualitative findings. On the technical side, low runtime overhead indicated that the system could sustain responsive viewing during the exhibition, helping preserve the spatial continuity of artworks as visitors moved through the site \cite{pico_pxrc}. On the curatorial side, participants’ accounts suggested that coherence also depended on how works were sequenced across zones and encountered along the circulation path, rather than on the stability of individual artworks alone.

Taken together, these findings highlight the need to coordinate runtime alignment with exhibition curation in MR exhibition deployment


\pb{Take-home message:} A pipeline for MR exhibition deployment should not be treated as a technical workflow alone, but as a coordinated process in which technical preparation and exhibition curation work together to achieve coherent exhibition realization.


\subsection{Pipeline's Effects on Artwork Interpretation and Exhibition Experience (RQ3)}

The proposed pipeline reshaped the relationship among artworks, environment, and viewers by enabling digital works to be experienced as situated parts of the exhibition space rather than detached overlays. The formal study results suggest that stronger spatial integration changed how viewers understood artworks in relation to their physical surroundings. Participants did not simply judge whether digital content was visible at a location; rather, they distinguished between works that felt meaningfully embedded in the site and works that appeared as detached overlays. Works were perceived most positively when they responded to their surroundings in visual, spatial, or semantic terms, whereas poorly integrated works were more likely to be described as arbitrary and interchangeable with other locations. In this sense, the pipeline enabled artworks to be experienced as belonging to the exhibition site, rather than merely appearing within it.

This shift also changed how artworks were interpreted and remembered. Participants repeatedly indicated that when a work was well integrated into its surroundings, the content and its location were remembered together, suggesting that spatial context became part of the exhibition experience itself. Some participants further described the interaction between artwork and environment as producing something new beyond the artwork itself, implying that the site became part of the work's expressive condition rather than remaining a neutral backdrop. Importantly, participants did not attribute this effect to the system autonomously ``understanding'' space. Instead, they understood spatial fit as the result of curatorial and design intent, made possible through the proposed pipeline. In this sense, the pipeline not only supported viewers' interpretation, but also provided curators and designers with a practical means of translating site-specific intentions into exhibition experience. In conclusion, these findings suggest that the value of the proposed pipeline lies not only in stabilizing digital content in space, but in reshaping how artworks, site, and viewers are connected in MR exhibition experience.

\pb{Take-home Message:} The value of the proposed pipeline lies not only in stabilizing digital content in space, but in more tightly connect artworks and viewers together in MR exhibition experience.

\section{Conclusion and Limitation}

\pb{Summary \& implication:} In this paper, we propose a practical pipeline for large-scale MR exhibition deployment and evaluated it from both system and audience perspectives. Through a pilot study, we compared the commonly-used marker-based and SLAM-based aligning methods \cite{reviewmixed2020}, showing that how the choice of spatial alignment method would impact viewers' experience of continuity and immersion. Our findings highlight the effective role of spatial alignment in MR exhibition settings \cite{radanovic2023aligning}. Afterwards, we propose a pipeline with SLAM-based method and use it to deploy a large-scale MR exhibition in real world. We emphasize that the technical deployment and exhibition curation should be treated as a coordinated process. Corresponding users' feedback evidenced that such an coordination could support a higher-quality exhibition delivery, enhancing their on-site experience. Last but not least, through the qualitative analysis on users' responses, we found our pipeline also shaped the relationship between artworks and the spaces where they displayed, by fostering a feeling of connecting the artworks and exhibition environment as a whole.

In a nutshell, this work provides a practical reference and empirical understanding for future MR exhibition deployment. For exhibition curators and organizers, it offers a workable way to translate curatorial intentions into on-site exhibition structures by coordinating technical deployment with exhibition curation. For researchers and developers, it highlights that spatial alignment should be understood in relation to experiential outcomes, rather than merely as a technique for artwork placement.
And for artists, it opens new creative direction by extending MR from a medium of digital overlay to a site-responsive mode of deployment.

\pb{Limitation \& future work:} Several limitations remain in this work. The first one is the consideration about the representativeness of the participant group. The current audiences were drawn mainly from a campus population aged 18 to 35, many of whom were relatively familiar with digital devices. Future studies will address this limitation by examining the pipeline with broader user groups like the elders and children, exploring how diverse user groups react towards our pipeline. Another limitation is that our pipeline mainly provided basic supporting functions with limited interaction design towards the artworks. It also does not support interaction among the visitors. Thus, future work could incorporate richer interactive functions and examine their effects on the MR exhibition experience.

\clearpage
\bibliographystyle{ACM-Reference-Format}
\bibliography{01_Texts/reference}

\end{document}